\documentclass[11pt] {article}
\usepackage{times}
\usepackage{graphics}
\usepackage{tensor}
\usepackage{natbib}
\usepackage{makeidx}
\usepackage{latexsym}
\usepackage{bm}
\usepackage{amsmath}
\usepackage{amsthm}
\usepackage{amssymb}
\makeindex
\newcommand{\beq}{\begin{equation}}
\newcommand{\eeq}{\end{equation}}

\newcommand{\defn}{\begin{quote}{\bf Definition. }}
\newcommand{\edefn}{\end{quote}}
\newcommand{\thm}{\begin{theorem}}
\newcommand{\ethm}{\end{theorem}}

\newcommand{\bmat}[1]{\left ( \begin{array}{#1}}
\newcommand{\emat}{\end{array}\right )}
\theoremstyle{definition}

\theoremstyle{plain}
\newtheorem{theorem}{Theorem}

\newcommand{\eps}[3]
{{\begin{center}
 \rotatebox{#1}{\scalebox{#2}{\includegraphics{#3}}}
 \end{center}}
}

\begin{document}

\title{A note on basis dimension selection in generalized additive modelling} %\tnoteref{t1,t2}}
%\tnotetext[t1]{This document is a collaborative effort.}
%\tnotetext[t2]{The second title footnote which is a longer
%longer than the first one and with an intention to fill
%in up more than one line while formatting.}

\author{Natalya Pya and Simon N Wood\footnote{{\tt simon.wood@bath.edu}}\\ ~ \\ Nazarbayev University and University of Bristol}

\maketitle
\begin{abstract}

Two new approaches for checking the dimension of the basis functions when using penalized regression smoothers are presented. The first approach is a test for adequacy of the basis dimension based on an estimate of the residual variance calculated by differencing residuals that are neighbours according to the smooth covariates. The second approach is based on estimated degrees of freedom for a smooth of the model residuals with respect to the model covariates. In comparison with basis dimension selection algorithms based on smoothness selection criterion (GCV, AIC, REML) the above procedures are computationally efficient enough for routine use as part of model checking.

\end{abstract}

\section{Introduction}

Penalized regression smoothing splines have been used extensively for representation of smooth model terms in generalized additive modelling \citep{sil86,eil96,wood06a}. In comparison with full rank smoothing splines \citep{wah90,gu02}, with  a free parameter for each data point, penalized regression splines are computationally attractive as the basis dimension, $k$, is set to much less than the number of data points. Provided that $k$ is large  enough to avoid oversmoothing/underfitting, its exact value is rather unimportant, since overfitting and the effective degrees of freedom of the smooth are then controlled by the weight given to the smoothing penalty during estimation, which is determined by the smoothing parameter, $\lambda$, rather than $k$. However it is necessary to check that $k$ really is large enough, and here useful readily applied checking methods are lacking. 

The choice of $k$ has not been widely discussed in the literature. \cite{rup02} proposed two algorithms based on minimizing GCV over a set of specified basis dimensions. He demonstrated empirically that there is a minimal threshold for the number of knots needed for a good fit, larger numbers of knots have only slight influence on the fit. Theoretical justifications are presented in \cite{lir08} who also show that the threshold depends on the degree of the spline order. \cite{kau11} introduced a likelihood based criterion for selecting $k$ in the case of the mixed-model specification of the penalized spline \citep{rup03,wand08}. In this specification the smoothing parameter $\lambda$, is represented as the ratio of the residual variance and the variance of the spline coefficients which allows estimating $\lambda$ simultaneously with the coefficients by likelihood maximization. This is equivalent to the restricted maximum likelihood criterion,  REML \citep{wood06a}. $k$ is treated as an additional discrete parameter of the log likelihood in \cite{kau11}. Two nested optimizations are proposed for parameters estimation: for each value for $k$ log likelihood is re-maximized with respect to model parameters. 

Searching for the `optimal' $k$ is clearly a computationally expensive option, and arguably of limited practical utility given the empirical and theoretical evidence that all that really matters is that $k$ is not too small. This note therefore proposes two computationally efficient approaches for checking the adequacy of the basis dimension from a fitted model, and compares their practical efficacy to a fuller search for an optimal $k$, in a small simulation study.

\section{Methods for checking the basis dimension}
 We consider a generalized additive model
\beq
g(\mu_i) = {\bf A} {\bm \theta} + \sum_{j=1}^p f_j(x_{ji}),~~~ Y_i \sim {\rm  EF}(\mu_i,\phi), ~~~ i=1,\ldots,n,\label{gam}
\eeq
where $Y_i$ are independent univariate response variables from an exponential family distribution with mean $\mu_i$ and scale parameter $\phi$, $g$ is a known smooth monotonic link function, $\bf A$ is a model matrix, $\bm \theta$ is a vector of unknown parameters, $f_j$ is an unknown smooth function of predictor variable $x_j$ that may be vector valued.

For representing the smooth functions $f_j(x_j)$ various penalized regression smoothers are available, such as cubic regression splines and P-splines for smooths of a single covariate, or thin plate regression splines and tensor product smooths for representing smooths of several covariates. The idea is to specify a basis for each function and choose an appropriate set of basis functions, $B_{jt},$ so that the $j^{th}$ smooth function can be represented as
$$f_{j}(x_{j})=\sum_{t=1}^{k_{j}}B_{jt}(x_{j})\beta_{jt},$$
where $\beta_{jt}$ are coefficients to be estimated, and $k_{j}$ is a number of basis functions.
After selecting smoothing penalties for each smooth function, the penalized log likelihood function
is maximized to get estimates of $\beta_{jt}$ given smoothing parameters. The smoothing parameters, $\lambda_j$, controlling the strength of penalization can be estimated by optimizing criteria such as GCV, AIC, or REML.

Although the smoothness  of each smooth term in (\ref{gam}) is controlled by the smoothing parameter, $\lambda_j$, the basis dimension $k_j$ used for smooth representation needs to be sufficiently large to avoid oversmoothing.
To ensure that $k_j$ is adequate the following approaches will be considered
\begin{enumerate}
  \item A hypothesis testing approach based on comparison of the residual variance estimate from the whole model fit, with the residual variance estimated by differencing residuals that are `neighbours' according to the covariates of the smooth term under consideration. 
  \item A procedure based on re-smoothing the model residuals with respect to the covariates of the smooth of interest, using a higher basis dimension, in order to search for missed pattern in the residuals. 
  \item Optimizing the GCV or REML criterion with respect to the basis dimensions, as well as the smoothing parameters.
\end{enumerate}
The first two procedures do not seem to have appeared in the literature, while the third is essentially what is proposed in \cite{rup02} and \cite{kau11}. Option 1 is simple and efficient enough for routine automatic use.

\subsection{Hypothesis testing approach \label{k-check}}

To test whether $k_j$ is sufficiently large for representing $f_j(x_j)$ the following test is proposed.
Let $r_i$ denote deviance residuals for the model that includes $f_j(x_j)$, in the Gaussian case simply $r_i=y_i-\hat{\mu}_i$. 
First consider a smooth of a univariate $x_j$. Let $\tilde {\bf r}$ denote the vector of residuals $r_i$, ordered according to the values of the observed covariates, $x_{ji}$. Define $\Delta_i = \tilde r_{i+1} - \tilde r_i$ for $i = 1, \ldots, n-1$. Then $\phi_\Delta = \frac{1}{2n-2} \sum_{i=1}^{n-1} \Delta_i^2$ is an estimate of the scale parameter, and if $\hat \phi$ is the corresponding estimate from the model fit, we can define a statistic 
$$
\kappa = \phi_\Delta/\hat \phi
$$
which should be close to one if the basis dimension is adequate, but should be less than one if $k_j$ is too small so that there is residual pattern left in the residuals. 

In the case of multivariate $x_j$, then the $M$ nearest neighbours of each observation can be found based on the proximity of $x_j$ vectors. This has relatively modest total cost $O(nM\log(n))$, see Press et al. (2007). Let $m_{ij}$ be the index of the $j{\rm th}$ nearest neighbour of observation $i$ (excluding self). Then we define 
$
\Delta_{ij} = r_i - r_{m_{ij}}
$ for $i = 1, \ldots n$, $j = 1, \ldots, M$, and $\phi_\Delta = \frac{1}{2Mn} \sum_{j=1}^M\sum_{i=1}^{n} \Delta_{ij}^2$. The expression for $\kappa$ is as before. 

The distribution of $\kappa$ under the null hypothesis that $k_j$ is adequate (so that there is no pattern in the residuals ordered with respect to the covariate), can be simulated by repeatedly randomly re-shuffling the original vector of residuals, and re-computing $\kappa$ for each replicate. A p-value can then be obtained. If it is too small, we should consider increasing $k_j$.

\subsection{Residual re-smoothing }

A second simple method for checking $k_j$ is as follows.
\begin{enumerate}
  \item Fit the model with your choice of $k_j$ and extract the deviance residuals, $r_i$, $i=1,\ldots, n$.
  \item Create a smooth, $f^*_j(x_j)$, equivalent to $f_j(x_j)$, but with basis dimension $2k_j$, then estimate the model 
  $$
    r_i=f_j^*(x_{ji})+e_i, 
  $$
  \item If the estimate of $f^*_j$ indicates that there is pattern in the residuals (e.g. if the effective degrees of freedom of $\hat f^*$ is more than the minimum possible for such a smooth) then $k_j$ may not be large enough, and an increase of $k_j$ should be considered.
\end{enumerate}

It is also worth considering changes in the smoothing selection criterion after fitting with increased $k_j$. If the value of the criterion is increased or the decrease is not more than $2\%$ of the previous value, then it is suggested to stop doubling $k_j$ since it is not expected to improve the criterion \citep{rup02}.

\subsection{GCV/REML based approach}

The final approach is the simplest,  but also the least computationally efficient. The idea is simply to search over a specified set  $k_j$ values for the optimizer of the criterion used for smoothing parameter estimation. The full model has to  be re-fitted for each set of  trial values of $k_j$, so this is quite costly computationally. Such algorithms were proposed first by \cite{rup00} and \cite{rup02} for P-splines with the basis dimension selection by minimizing the GCV criterion, while \cite{kau11} suggests a similar likelihood based method.

\section{Illustrative simulated examples}

\begin{figure}
\begin{center}
\eps{-90}{.4}{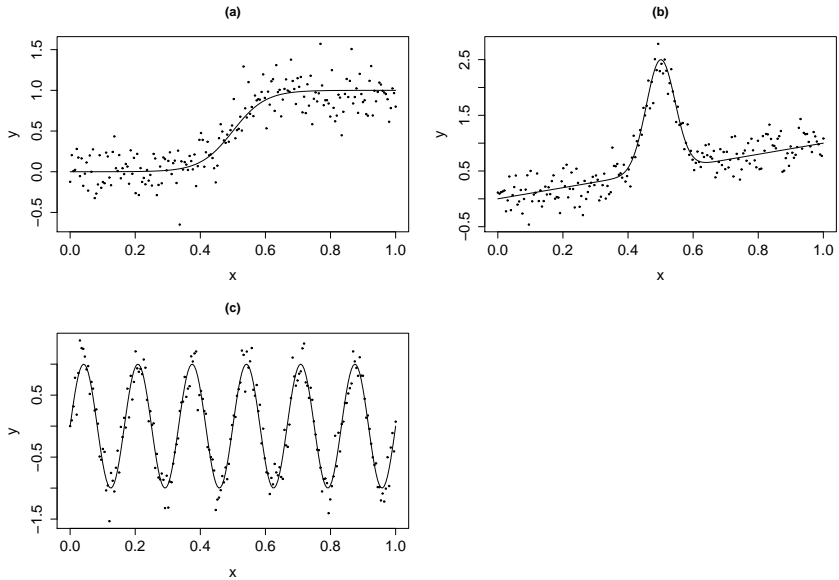}
 \end{center}
 \caption{Shapes of the univariate test functions used for simulation study.}
 \label{fig_uni_shapes}
\end{figure}

\begin{figure}
\begin{center}
\eps{-90}{.4}{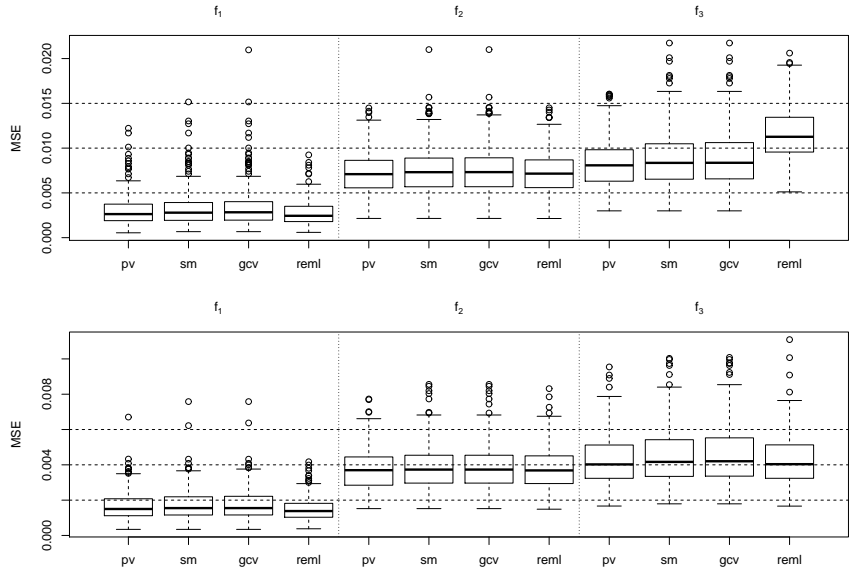}
 \end{center}
 \caption{MSE comparison between hypothesis testing approach (\texttt{pv}), residuals smoothing (\texttt{sm}), GCV search (\texttt{gcv}), and REML search (\texttt{reml}) methods for three single univariate smooth term models. The upper panel shows the results for $n=100$, the lower for $n=200$.}
 \label{fig_3uni_mse}
\end{figure}

\begin{figure}
\begin{center}
    \eps{-90}{.4}{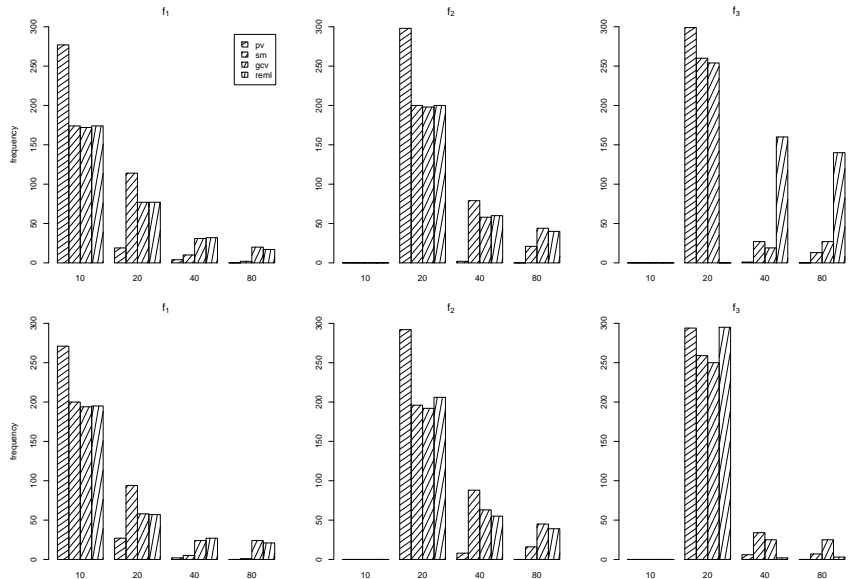}
 \end{center}
 \caption{Bar plots of $k$ selected by the hypothesis testing approach (\texttt{pv}), residuals smoothing (\texttt{sm}), GCV search (\texttt{gcv}), and REML search (\texttt{reml}) methods for three single univariate smooth term models. Three upper panels show the results for $n=100$, the lower panels for $n=200$.}
 \label{fig_3uni_k}
\end{figure}

The approaches were compared on five simulated examples of Gaussian data: three univariate single smooth models following \cite{rup02}, a bivariate single smooth model and an additive model with two smooth terms.
$$
y_i = f_t(x_i) + e_i, ~~ t=1,2,3,~~~ y_i = f(x_{1i},x_{2i}) + e_i, ~~~ y_i = f_1(x_{1i}) + f_2(x_{2i}) +e_i, ~~~ e_i \sim \textrm{N}(0,~\sigma^2).
$$
For all examples data were simulated independently from $y_i \sim \textrm{N}(\mu_{i},~0.2^2)$.
For univariate cases $x_i$ were equally spaced on $[0,1].$ Three test functions $f_t(x_i)$ with shapes shown in Fig. \ref{fig_uni_shapes} were applied.
 For sample sizes $n=100$ and $200$, $300$ replicates were simulated. Each single smooth regression model was fit using the R package \verb"mgcv" \citep{wood06a}. Thin plate regression splines with the second order derivative in penalty were used for representation of all smooths \citep{wood03}. The basis dimension was selected by the approaches presented in section 2. For methods 1 and 2 the basis dimension for a term was doubled if the check suggested that the current dimension was inadequate. Two variants of method 3 were tried, one using GCV and the other REML. The tested values of $k$ were 10, 20, 40, 80. The performance of the all methods was evaluated by the mean sum of squared differences between the fitted, $\hat{f}(x),$ and true values of $f_t(x),$
$$
  \textrm{MSE}=n^{-1}\sum_{i=1}^n\left\{\hat{f}(x_i)-f_t(x_i)\right\}^2.
$$
The results are summarized in Fig. \ref{fig_3uni_mse}. Fig. \ref{fig_3uni_k} shows bar plots of $k$ selected by the four algorithms.
The simulation results show that the performance of all four algorithms is quite similar excepting REML when fitting the sine function with $n=100$. In that case REML tended to select a larger basis dimension than other methods which resulted in larger MSE.
The basis dimensions required for fitting the $f_1$ are around 10 -- 20 which is typical for many monotone functions used in practice. It is clear that $k=10$ is too small for the `bump' function $f_2$ and sine waves with six cycles $f_3$. All algorithms selected at least 20 basis functions to give satisfactory fits for those much `wigglier' functions. It can be noticed that the hypothesis testing approach stops earlier and chooses smaller basis dimensions compared to the other methods.

\begin{figure}
\begin{center}
\eps{-90}{.4}{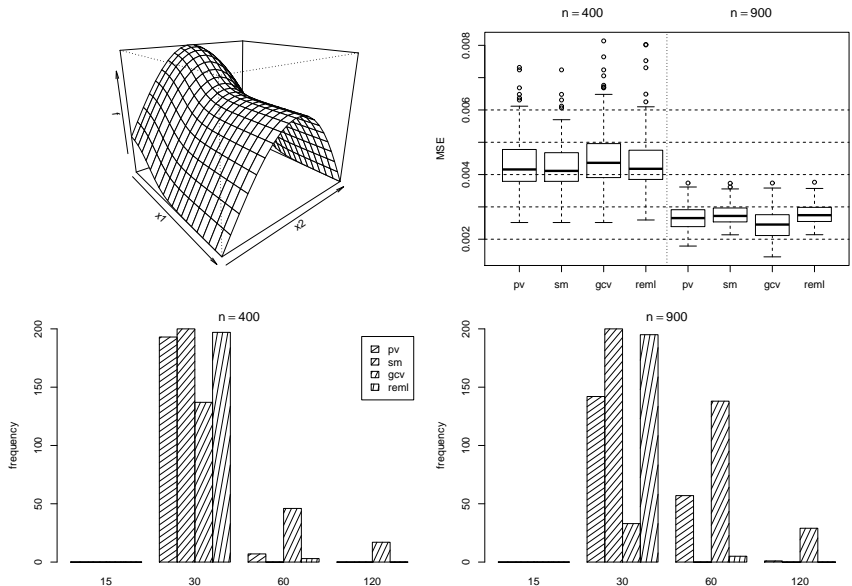}
 \end{center}
 \caption{Plots for bivariate example. Upper left: perspective plot of the bivariate test function; upper right: MSE comparison between four approaches; lower left: bar plots of $k$ selected for $n=400$; lower right: bar plots of $k$ selected for $n=900$.}
 \label{fig_bivar}
\end{figure}

The shape of the bivariate test function used in the next example is shown in Fig. \ref{fig_bivar}. 20 and 30 values of each of the two covariates, $x_{1i}$ and $x_{2i}$, were equidistant in $[-1,3]$ and $[0,1]$, giving samples of sizes $n=400$ and $900$. 200 replicate data sets were simulated for each sample size. The trial values of the basis dimension were 15, 30, 60, and 120. While setting $k=15$ is too low for all the approaches, $k=30$ appeared to be enough with only GCV choosing occasionally larger $k$ for $n=400$ and more often for $n=900$ (see Fig. \ref{fig_bivar}).

\begin{figure}
\begin{center}
\eps{-90}{.4}{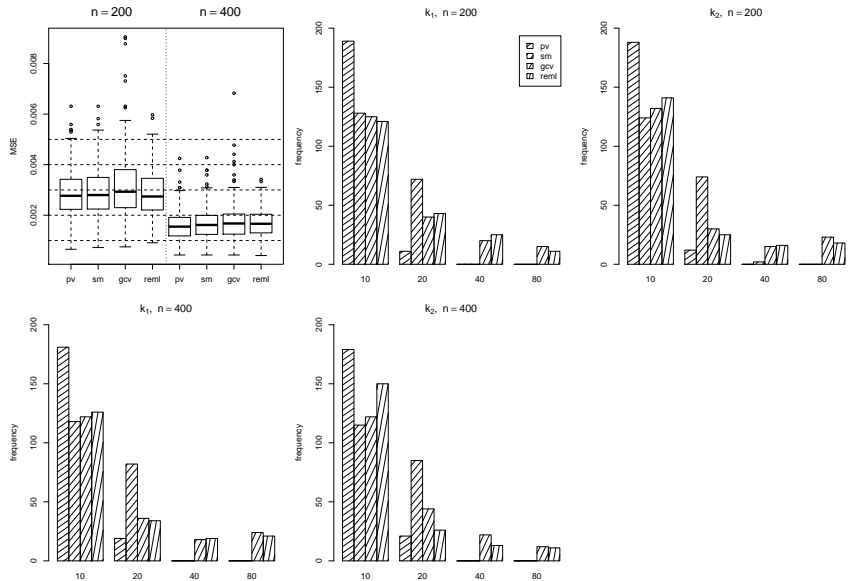}
 \end{center}
 \caption{Plots for the additive model. Upper left: boxplots of the MSE of four approaches for $n=200$ and $400$; upper middle and right: bar plots of $k_1$ and $k_2$ selected for two smooth terms, $f_1$ and $f_2,$ for $n=200$; lower panels: bar plots of $k_i$ for two smooths for $n=400$.}
 \label{fig_addit}
\end{figure}

The last example is an additive model with two smooth terms, $y_i = f_1(x_{1i}) + f_2(x_{2i}) +e_i,$ where $f_1$ is the first function used in the univariate case and $f_2$ is a sine function with a single cycle. $x_{1i}$ and $x_{2i}$ were i.i.d. $\textrm{U}(0,1)$. Sample sizes $n=200$ and $400$ were considered. The simulation results of this study based on 200 replicates are shown in Fig. \ref{fig_addit}.
The behaviour of all algorithms is similar to that of the univariate example. However, the main benefit of the hypothesis testing and re-smoothing residuals approaches is in their computational efficiency. The GCV/REML algorithms require the whole model to be re-fitted at every combination of $(k_1,k_2)$ values over the specified grid. But the hypothesis testing approach needs only calculating the estimate of the residual variance and its p-value. And the second approach requires re-smoothing of a single term rather than the whole model. Re-fitting the whole model with an increased value of the suspected $k_j$ occurs only in case of a low p-value or high edf for the particular smooth term. The computational advantage grows with the number of smooth terms in the GAM.

\section{Conclusions}

In practice the simple and efficient basis dimension check proposed in Section \ref{k-check} appears to perform as well as more expensive approaches requiring multiple model fits. Given its computational efficiency it seems sensible to incorporate this check as a standard part of model checking for penalized regression spline based generalized additive models. Of course as with any model checking, the methods have to be used with care. Mean variance relationship problems and residual autocorrelation problems unrelated to $k_j$ can obviously cause any of the methods considered here to suggest increasing $k_j$, when the real problem lies elsewhere. 

\bigskip

\noindent This work was supported by EPSRC grants EP/K005251/1 and EP/1000917/1

%\section*{Acknowledgments}

\bibliographystyle{plainnat}
%\bibliographystyle{apalike}
%\bibliographystyle{num-names}

%\nocite{*}

\bibliography{chooseKbib}

\end{document}